# Fabrication and Characterization of the Moiré surface state on a topological insulator

*Yi Zhang[1], Dang Liu[1], Qiaoyan Yu[1], Ruijun Xi[1], Xingsen Chen[1], Shasha Xue[1], Jice Sun[1], Xian Du[1], Xuhui Ning[1], Tingwen Miao[1], Pengyu Hu[1], Hao Yang[1,2,3], Dandan Guan[1,2,3], Xiaoxue Liu[1,2,3], Liang Liu[1,2,3], Yaoyi Li[1,2,3], Shiyong Wang[1,2,3], Canhua Liu[1,2,3], Haijiao Ji[1], Noah F. Q. Yuan [1], Hao Zheng[1,2,3] *, and Jinfeng Jia[1,2,3,4,5] **

1, Tsung-Dao Lee Institute, Key Laboratory of Artificial Structures and Quantum Control (Ministry of Education), School of Physics and Astronomy, Shanghai Jiao Tong University, Shanghai 200240, China

2, Hefei National Laboratory, Hefei 230088, China

3, Shanghai Research Center for Quantum Sciences，99 Xiupu Road，Shanghai 201315, China

4, Department of Physics, Southern University of Science and Technology, Shenzhen 518055, China

5, Quantum Science Center of Guangdong-Hong Kong-Macao Greater Bay Area (Guangdong), Shenzhen 518045, China

*Corresponding author. E-mail: haozheng1@sjtu.edu.cn ; jfjia@sjtu.edu.cn

Abstract:

A Moiré superlattice on the topological insulator surface is predicted to exhibit many novel properties but has not been experimentally realized. Here, we developed a two-step growth method to successfully fabricate a topological insulator $Sb_2Te_3$ thin film with a Moiré superlattice, which is generated by a twist of the topmost layer via molecular beam epitaxy. The established Moiré topological surface state is characterized by scanning tunneling microscopy and spectroscopy. By application of a magnetic field, new features in Landau levels arise on the Moiré region compared to the pristine surface of $Sb_2Te_3$, which makes the system a promising platform for pursuing next-generation electronics. Notably, the growth method, which circumvents contamination and the induced interface defects in the manual fabrication method, can be widely applied to other van der Waals materials for fabricating Moiré superlattices.

In recent years, Moiré material has received widespread attention, especially the twisted bilayer graphene and transition-metal dichalcogenides.[1-5] Due to the interlayer coupling modulated by the twist angle, the band structure of the Moiré materials is extensively reconstructed. A flat band arises and leads to various emergent quantum phases, including correlated insulating phases,[2,6] superconductivity,[3] orbital magnetism,[7] quantum and fractional quantum anomalous Hall states,[8-10] Moiré excitons,[11,12] generalized Wigner crystal states,[13] and so on. Meanwhile, theorists predict that Moiré superlattices will also give rise to many intriguing phenomena in topological insulators (TIs). For example, the Moiré potential may renormalize the Fermi velocity of the original surface Dirac cone and create additional satellite Dirac cones.[14] High-order van Hove singularities and its power-law divergent density-of-states-enhanced superconductivity may appear on Moiré topological surface states.[15] Being subjected to a high magnetic field, Landau quantization of the electrons in Moiré topological surface states also exhibits unusual behaviors.[16] However, there are major differences between TIs and two-dimensional (2D) materials. 2D Moiré materials can be fabricated by manually stacking two mechanically exfoliated atomic layers with a certain twist angle. However, surface states in 3D TIs arise only with thicknesses of the material exceeding a certain value, i.e., the critical thickness. Therefore, attaching a monolayer of the same material with a twist angle or a monolayer of other insulating material is believed to be the (only) practical way. Unfortunately, unlike graphene, TIs such as $Sb_2Te_3$ feature stronger interlayer interactions and enhanced brittleness,[17] making it challenging to exfoliate one layer of TI (free of defects) and attach it to another TI bulk. To date, building Moiré surface states in TIs still remains elusive, despite some devoted efforts.[18]

In this work, a $Sb_2Te_3$ film with a Moiré pattern on the top surface was successfully grown by molecular beam epitaxy (MBE). The Moiré pattern was achieved by twisting the top layer of $Sb_2Te_3$ through a two-step growth method.[19] The influence of the Moiré pattern on Dirac cones, band structures, and local electron density of states had been systematically studied. With an applied magnetic field of up to 10 T, Landau levels were established in the samples. The two-Dirac-cone model was found to agree with the Landau level on the Moiré topological surface state for various Moiré superlattice periods.

The experiments were conducted in an MBE-scanning tunneling microscopy (STM) system with a base pressure of < 1 × 10-10 Torr. An n-type Si(111) 7 × 7 surface was used as the substrate. $Sb_2Te_3$ films were grown by the evaporation of high-purity Sb and Te (both 99.9999%) from Knudsen cells. The temperatures of Te and Sb sources were set at 230 and 280 °C, respectively. The substrate temperature was maintained at 200-220 °C for the first 35 min of growth and then increased to 250 °C from the 35th to 50th minute. After growth, the sample was transferred *in situ* to a scanning tunneling microscope at 4.5 K. d$I$/d$V$ spectra were obtained by a lock-in technique with a bias modulation of 1 mV at 931 Hz.

The $Sb_2Te_3$ class of crystals, including $Bi_2Se_3$ and $Bi_2Te_3$, is a prototypical TI. Five atomic sheets are covalently bonded in the order Te-Sb-Te-Sb-Te to form 1 quintuple layer (QL) with a thickness about 1 nm, while the interaction between adjacent QLs is van der Waals (vdW) force. Previous studies found that only when a $Sb_2Te_3$ film is thicker than 4 QLs is a topological surface state, which is a single spin-momentumlocked Dirac cone at the Γ point, established.[20,21] We grew $Sb_2Te_3$ thin films on Si substrates with a designed two-step growth method, as illustrated in Figure 1a. At the initial state of growth, $Sb_2Te_3$ nucleates into a large number of 1-QL high islands with various rotational angles compared to the Si lattice. Figure 1b shows that two such islands with angles of $\theta 1$ and $\theta 2$ merge together, which leads to a thin film with two rotational domains. On the basis of the atomically resolved STM images in different domains shown in Figure 1c,d, we can observe that the two domains are rotated about 12.8° to each other, i.e., $\theta 1$ - $\theta 2$ = 12.8°. One may note that a Moiré pattern already emerges in the 1-QL film (shown in Figure 1b), which is induced by lattice mismatch between $Sb_2Te_3$ and Si(111).[22] However, such a Moiré pattern disappears due to stress relaxation as the film thickens, such as the 4-QL film in Figure 1e, but domain boundaries remain, which demonstrates that the rotational domains persist during growth.

As demonstrated in Figure 1a, if one is able to make the top layer of one domain extend to another, a Moiré superlattice is consequently established due to the twist angle between the topmost layer and the underlying layer of $Sb_2Te_3$. This Moiré system is built by the twisted same material and thus should minimize lattice mismatch, strain, and interface defects, among other drawbacks that are easily found in heterojunctions.[23] We apply the two-step growth method to obtain such a Moiré pattern. Figure 2a represents one example. The $Sb_2Te_3$ thin film contains two domains rotated 11.0 ± 0.3° to each other, which are evidenced by the atomic images on each domain (Figure 2b,c). The eighth QL of $Sb_2Te_3$ on one domain expands to another and covers the underneath seventh QL, leading to the formation of a Moiré superlattice with a period of 2.24 nm (Figure 2d). According to the formula for calculating the twist angle of Moiré superlattices[1] $\lambda_m = a/\sqrt{2(1-\cos\theta)}$, a Moiré superlattice with a twist angle of 11.0 ± 0.3° is theoretically expected to exhibit a period of 2.21 ± 0.06 nm, which is in good agreement with the experimental measurement of the 2.24 nm Moiré period at $Sb_2Te_3$ (lattice constant = 0.425 nm). Although $Sb_2Te_3$ contains five atomic layers within 1 QL, unlike graphene, which has only one atomic layer, the dependence of the Moiré superlattice period on the twist angle remains consistent.

The mechanism of the two-step growth method to achieve Moiré superlattices is explained as follows. The first step is the low-temperature growth of thin films with rotation domains. We kept the substrate at 200-220 °C during the deposition of $Sb_2Te_3$, which enables nucleation and vertical growth of defect-free films while maintaining the initially formed domain structures. Subsequently, the second step is high temperature growth. About 250 °C substrate temperature enhances the diffusion of the atoms, disrupts inter-QL vdW interactions while preserving in-plane covalent bonds, and thus leads to a lateral growth mode. The topmost layer of the material grows across the domain boundary and forms Moiré superlattices. We believe that this MBE growth method can be applied to a wide range of vdW materials.

We note that the height of 1 QL of $Sb_2Te_3$ in the Moiré region is slightly higher than that of the normal region. Moreover, the spectra are also different between the two regions. Namely, we find that the surface Dirac point shifts from 85 to 121 meV due to the Moiré pattern, as displayed in Figure 2e,f. This indicates that the Moiré pattern indeed modifies the electronic property of our samples, thus inspiring us to conduct more characterizations on the Moiré topological surface state.

A strong magnetic field can renormalize the electronic state in a crystal to Landau levels. Landau levels in a Moiré material are expected to display many unusual phenomena.[16,21] Parts a and b of Figure 3 show our d$I$/d$V$ spectra with the background subtracted at different magnetic fields (see the raw data in Figure S2) measured in a normal region, i.e., without a Moiré pattern, in the 8-QL film shown in Figure 2a. Landau levels are clearly resolved, exhibiting typical characteristics of Landau levels from the topological surface states of $Sb_2Te_3$. Data fitting yields a Fermi velocity of 3.43 × 10^5 m/s, which is consistent with previous reports.[21] We note that the negative branches of Landau levels are invisible in the data, which may be caused by the electrostatic potential introduced by the STM tip.[24] Regardless, it proves the high quality of our sample. We now turn our attention to the Moiré region. The d$I$/d$V$ spectra under magnetic fields in Figure 3c,d clearly reveal the establishment of Landau levels on a Moiré superlattice. By comparing parts a and c of Figure 3, one may notice that the Landau levels in the Moiré region differ from those in the normal region at high magnetic field. Namely, more peaks, which are indicated by the arrows in the d$I$/d$V$ spectrum at 10.5 T in Figure 3c, emerge and exhibit the characteristics of Landau-level splitting behavior. These additional peaks start to appear at a magnetic field of 6.3 T and become clearly visible under large magnetic fields. The reason for the Moire-induced new Landau-level features appearing at high field intensity is related to the size of the Moiré area. The spectra in Figure 3c were acquired on the sample shown in Figure 2a, where the effective radius of the Moiré pattern region is about 19 nm. On the basis of the formula of the cyclotron orbital radius of the $k$th Landau level[27] $r_k = \sqrt{(2k+1)\hbar/eB}$, we estimate that a magnetic field larger than 6.3 T starts to complete a cyclotron trace to establish the first Landau level and additional peaks begin to appear. To ensure that these additional peaks arise solely from the Moiré superlattice rather than charged defects, which have been reported to split the Landau levels in $Sb_2Te_3$, with such split peaks showing clear positional evolutions,[25,26] we measured two position-dependent d$I$/d$V$ spectra along the horizontal and vertical directions (Figure S3). Both spectra are spatially uniform, confirming that the observed new peaks in the Landau levels indeed originate from the Moiré superlattice.

To better investigate the changes in the electronic properties on the Moiré topological surface states, we invoke recent theories that have predicted that the Moiré potential on a TI can generate multiple Dirac cones in addition to the original one.[14,15] In Figure 4, we attempt to fit our Landau-level data, which are acquired on different samples with various twist angles and Moiré periods, with a two-Dirac-cone model. The first Dirac cone is the original TI surface state, whose Hamiltonian and Landau levels are

$$H_1 = v_1(\sigma_x \pi_y - \sigma_y \pi_x) \quad (1)$$

$$E_{n,1} = E_{D1} + \text{sgn}(n) v_{F1}\sqrt{2e\hbar|n|B} \quad n = 0, \pm1, \pm2, \ldots \quad (2)$$

The second Dirac cone' Landau quantization follows[25,28]:

$$H_2 = v_2(\sigma_x \pi_y - \sigma_y \pi_x) + gB\sigma_z \quad (3)$$

$$E_{n,2} = E_{D2} + \text{sgn}(n)\sqrt{2v_{F2}^2 e\hbar|n|B + g^2 B^2} \quad n = \pm1, \pm2, \ldots \quad (4)$$

$$E_{0,2} = E_{D2} - gB \quad n = 0$$

Here, $\sigma$ denotes the Pauli matrixes and $\pi$ denotes the canonical momentum. $E_{D1}$ and $E_{D2}$ are the energies of the first and second Dirac points, $v_{F1}$ ($v_{F2}$) is the Fermi velocity of the first (second) Dirac cone, and $g$ stands for the Landé $g$ factor, which is probably induced by the coupling between multiple Dirac points in the Moiré surface states.[28] We find that the two-Dirac-cone model agrees well with the data but the oneDirac-cone model clearly fails to fit all Landau-level peaks. The deduced parameters are listed in Table 1, from which one may note that the nonzero $g$ factor is essential to fitting the second Dirac cone, while the Zeeman effect is negligible in fitting the Landau levels of the first Dirac cone. In other words, the second Dirac cone induced by the Moiré superlattice potential exhibits a $g$ factor that is 1 order of magnitude greater than that of the intrinsic Dirac cone in the surface states of $Sb_2Te_3$. Consequently, compared to the normal region, the TI surface states in the Moiré superlattice region not only possess an additional Dirac cone but also have the zeroth Landau level of this Dirac cone, possessing a nonzero slope with respect to magnetic field variations. Such a nonzero $g$ factor holds potential for applications in the spin manipulation of topological surface states.[25] We believe our two-step growth method, which produced the Moiré topological surface states, together with future doping techniques, will exhibit more novel quantum properties.

In summary, we have developed a method to successfully realize the Moiré topological surface state in a twisted homojunction based on TI $Sb_2Te_3$. Spectroscopic characterizations reveal that the Moiré pattern induces new features in Landau levels, thereby proving that our method is effective for modifying the electronic properties of a TI. This work may open a new avenue for twistronics in topological nontrivial materials.

**Supporting Information.**

Topographic images of Moiré pattern under different bias voltages; raw data of Landau fans; position-dependent dI/dV spectra at Moiré region; position-dependent dI/dV spectra across the boundary; atomic images of the Moiré region boundary

**Author contributions:**

H. Z. and J.-F. J. conserved the project. Y. Z. conducted STM measurements and MBE growth, assisted by D. L., Q.-Y. Y., R.-J. X., X.-S. C., S.-S. X., J.-C. S., X. D., X.-H. N., T.-W. M., and P.-Y. H. All authors discussed the result and contributed to the paper writing.

**Acknowledgements:**

We thank NSFC (Grants No.12488101, No. 92365302, No. 12474156, No. 22325203, No. 52102336,No. 12474121), the Ministry of Science and Technology of China (Grants No. 2020YFA0309000，2024YFA1410100), the Science and Technology Commission of Shanghai Municipality (Grants No. 2019SHZDZX01, No. 20QA1405100, No.24LZ1401000)，Cultivation Project of Shanghai Research Center for Quantum Sciences (Grant No. LZPY2024-04) and innovation program for Quantum Science and Technology (Grant No. 2021ZD0302500) for financial support.

Figures:

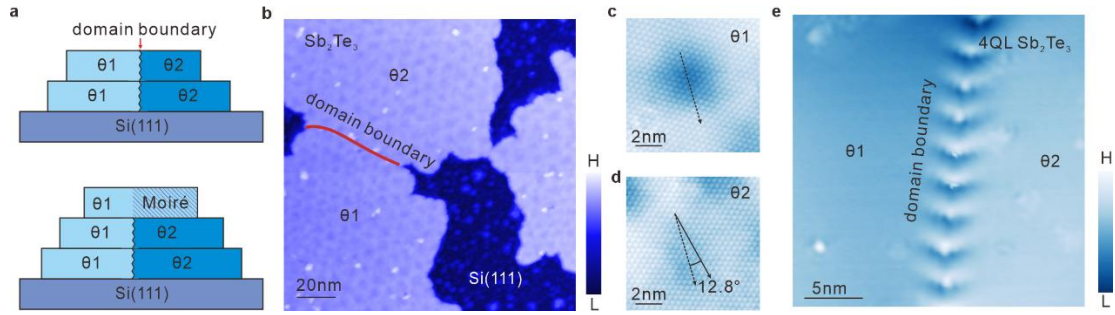

**Figure 1.** Rotation domains and a Moiré pattern on a $Sb_2Te_3$ thin film. (a) Schematic diagram of Moiré pattern formation. Upper panel: Two islands of $Sb_2Te_3$ were grown with angles of $\theta1$ and $\theta2$ with respect to the Si substrate. When they merge together, a thin film with rotation domains and a domain boundary arises. Lower panel: While the growth temperature increases, the top layer of one domain extends to another and forms a Moiré pattern. (b) Topographic image of 1-QL of $Sb_2Te_3$ on Si (*V*bias = 1 V; *I* = 100 pA). A Moiré pattern appears due to the lattice mismatch between $Sb_2Te_3$ and Si(111). $\theta1$ and $\theta2$ indicate the two rotation domains of the $Sb_2Te_3$ film. A red line marks the domain boundary. (c and d) Atomically resolved topographic images of the $Sb_2Te_3$ surface with atomic orientation angles $\theta1$ and $\theta2$ (*V*bias = -1 V; *I* = -100 pA), respectively. The two domains differ in angle by 12.8°. (e) Topographic image of 4-QL $Sb_2Te_3$. The Moiré pattern due to lattice mismatch disappears, but the domain boundary remains.

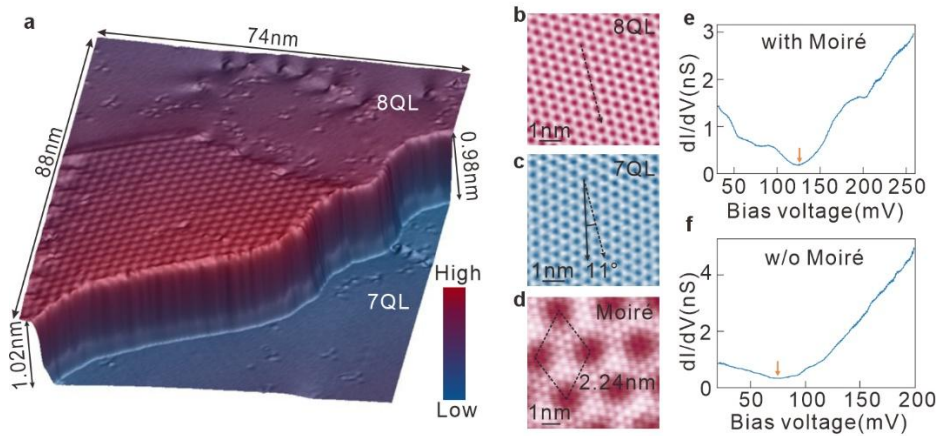

**Figure 2.** Moiré pattern due to the twisted top layer. (a) Topographic image of a $Sb_2Te_3$ thin film with two rotation domains and a Moiré pattern (*V*bias = 1 V; *I* = 100 pA). The upper and lower terraces are 8 and 7 QLs in height, respectively. (b and c) Atomic lattices of the normal regions at 8 and 7 QLs, respectively. The two domains differ by 11.0 ± 0.3°, resulting in a Moiré pattern with a period of 2.24 nm, whose atomic image is shown in part d. (e and f) d*I*/d*V* spectra measured on an 8-QL $Sb_2Te_3$ surface with and without a Moiré pattern. The arrows point to the Dirac points.

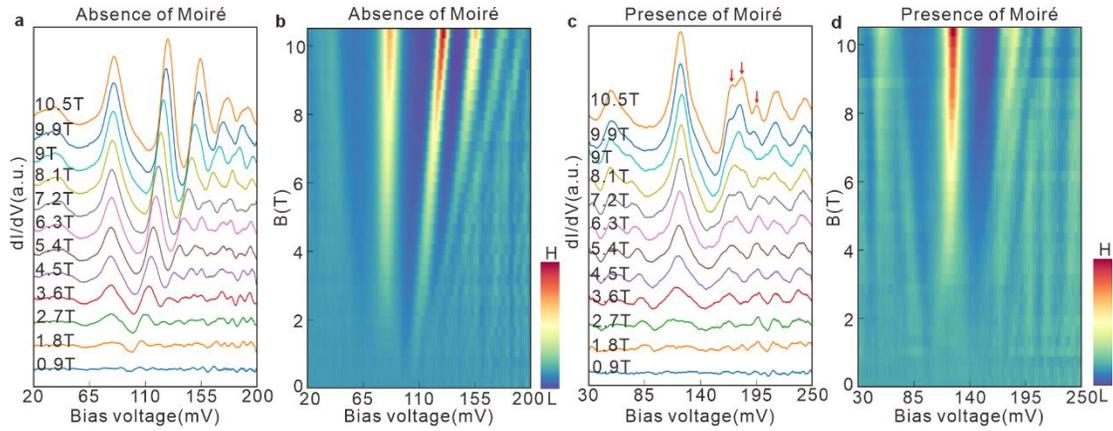

**Figure 3.** Figure 3. Moiré pattern that induced new features in the Landau level.

(a) d$I$/d$V$ spectra ($V$bias = 250 mV; $I$ = 300 pA) acquired on a normal region (e.g., without a Moiré pattern) of 8-QL $Sb_2Te_3$ (with the background subtracted). A magnetic field is applied from 0 to 10.5 T with an interval of 0.3 T. (b) Landau fan plot of the data in part a. (c) d$I$/d$V$ spectra ($V$bias = 250 mV; $I$ = 300 pA) measured on a Moiré region of 8-QL $Sb_2Te_3$ (with the background subtracted). Arrows point to the new features when compared to the data in the normal region. (d) Landau fan plot based on the data in part c.

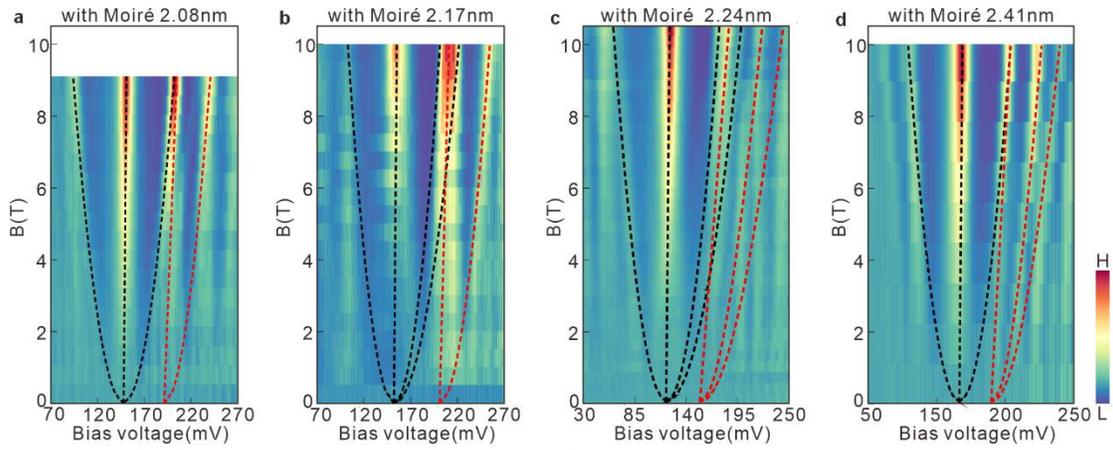

**Figure 4.** Landau fan fitting results under different Moiré periods.

Parts a-d show the Landau fans measured on four different Moiré regions with periods of 2.08, 2.17, 2.24, and 2.41 nm, respectively. A two-Dirac-cone model fits well with the data, and the red and black curves represent original and satellite Dirac cones, respectively.

|  | **2.08nm** | **2.17nm** | **2.24nm** | **2.41nm** |
|---|---|---|---|---|
| $E_{D1}$(meV) | 160 | 156 | 120 | 153 |
| $v_{F1}$(m/s) | $5.51 \times 10^5$ | $4.66 \times 10^5$ | $4.88 \times 10^5$ | $5.00 \times 10^5$ |
| $E_{D2}$(meV) | 211 | 209 | 155 | 189 |
| $v_{F2}$(m/s) | $4.85 \times 10^5$ | $5.00 \times 10^5$ | $5.09 \times 10^5$ | $4.50 \times 10^5$ |
| $g$ (meV/T) | 2.23 | 1.06 | 8.50 | 4.25 |
| $E_{D2} - E_{D1}$(meV) | 61 | 53 | 35 | 36 |

**Table1.** Table Title: The parameters used to fit Landau Levels on Four Moiré Regions

Two Dirac cone model was used to fit the Landau levels in four Moiré regions whose periods are listed in the first row. Energies of Dirac point ($E_{D1}$ and $E_{D2}$), Fermi velocities ($v_{F1}$ and $v_{F2}$) and g-factor are displayed.

**References:**

(1) Bistritzer, R.; MacDonald, A. H. Moiré Bands in Twisted Double-Layer Graphene. *Proc. Natl. Acad. Sci. U.S.A.* 2011, 108 (30), 12233–12237.

(2) Cao, Y.; Fatemi, V.; Demir, A.; Fang, S.; Tomarken, S. L.; Luo, J. Y.; Jarillo-Herrero, P. et al. Correlated Insulator Behaviour at Half-Filling in Magic-Angle Graphene Superlattices. *Nature* 2018, 556 (7699), 80–84.

(3) Cao, Y.; Fatemi, V.; Fang, S.; Watanabe, K.; Taniguchi, T.; Kaxiras, E.; Jarillo-Herrero, P. Unconventional Superconductivity in Magic-Angle Graphene Superlattices. *Nature* 2018, 556 (7699), 43–50.

(4) Andrei, E. Y.; MacDonald, A. H. Graphene Bilayers with a Twist. *Nat. Mater.* 2020, 19 (12), 1265–1275.

(5) Zheng, C.; Liu, X. Superconductivity and Topological Quantum States in Two-Dimensional Moiré Superlattices. *Quantum Front.* 2024, 3, 17.

(6) Chen, G. Correlated and Topological Physics in ABC-Trilayer Graphene Moiré Superlattices. *Quantum Front.* 2022, 1, 8.

(7) Lu, X.; Stepanov, P.; Yang, W.; Xie, M.; Aamir, M. A.; Das, I.; Efetov, D. K. et al. Superconductors, Orbital Magnets and Correlated States in Magic-Angle Bilayer Graphene. *Nature* 2019, 574 (7780), 653–657.

(8) Serlin, M.; Tschirhart, C. L.; Polshyn, H.; Zhang, Y.; Zhu, J.; Watanabe, K.; Young, A. F. et al. Intrinsic Quantized Anomalous Hall Effect in a Moiré Heterostructure. *Science* 2020, 367 (6480), 900–903.

(9) Lu, Z.; Han, T.; Yao, Y.; Reddy, A. P.; Yang, J.; Seo, J.; Ju, L. et al. Fractional Quantum Anomalous Hall Effect in Multilayer Graphene. *Nature* 2024, 626 (8000), 759–764.

(10) Cai, J.; Anderson, E.; Wang, C.; Zhang, X.; Liu, X.; Holtzmann, W.; Xu, X. et al. Signatures of Fractional Quantum Anomalous Hall States in Twisted $MoTe_2$. *Nature* 2023, 622 (7981), 63–68.

(11) Jin, C.; Regan, E. C.; Yan, A.; Iqbal Bakti Utama, M.; Wang, D.; Zhao, S.; Wang, F. et al. Observation of Moiré Excitons in $WSe_2$/$WS_2$ Heterostructure Superlattices. *Nature* 2019, 567 (7746), 76–80.

(12) Kai, F.; Wang, X.; Xie, Y. et al. Distinct Moiré Exciton Dynamics in $WS_2$/$WSe_2$ Heterostructure. *Quantum Front.* 2025, 4, 2.

(13) Regan, E. C.; Wang, D.; Jin, C.; Bakti Utama, M. I.; Gao, B.; Wei, X.; Wang, F. et al. Mott and Generalized

Wigner Crystal States in WSe$_2$/WS$_2$ Moiré Superlattices. *Nature* 2020, 579 (7799), 359–363.

(14) Cano, J.; Fang, S.; Pixley, J. H.; Wilson, J. H. Moiré Superlattice on the Surface of a Topological Insulator. *Phys. Rev. B* 2021, 103 (15), 155157.

(15) Wang, T.; Yuan, N. F.; Fu, L. Moiré Surface States and Enhanced Superconductivity in Topological Insulators. *Phys. Rev. X* 2021, 11 (2), 021024.

(16) Paul, N.; Crowley, P. J.; Devakul, T.; Fu, L. Moiré Landau Fans and Magic Zeros. *Phys. Rev. Lett.* 2022, 129 (11), 116804.

(17) Hou, F.; Yao, Q.; Zhou, C. S.; Ma, X. M.; Han, M.; Hao, Y. J.; Lin, J. et al. Te-Vacancy-Induced Surface Collapse and Reconstruction in Antiferromagnetic Topological Insulator MnBi$_2$Te$_4$. *ACS Nano* 2020, 14 (9), 11262–11272.

(18) Schouteden, K.; Li, Z.; Chen, T.; Song, F.; Partoens, B.; Van Haesendonck, C.; Park, K. Moiré Superlattices at the Topological Insulator Bi$_2$Te$_3$. *Sci. Rep.* 2016, 6, 20278.

(19) Guha, P.; Park, J. Y.; Jo, J.; Chang, Y.; Bae, H.; Saroj, R. K.; Yi, G. C. et al. Molecular Beam Epitaxial Growth of Sb$_2$Te$_3$–Bi$_2$Te$_3$ Lateral Heterostructures. *2D Mater.* 2022, 9 (2), 025006.

(20) Jiang, Y.; Sun, Y. Y.; Chen, M.; Wang, Y.; Li, Z.; Song, C.; Zhang, S. B. et al. Fermi-Level Tuning of Epitaxial Sb$_2$Te$_3$ Thin Films on Graphene by Regulating Intrinsic Defects and Substrate Transfer Doping. *Phys. Rev. ett.* 2012, 108 (6), 066809.

(21) Jiang, Y.; Wang, Y.; Chen, M.; Li, Z.; Song, C.; He, K.; Xue, Q. K. et al. Landau Quantization and the Thickness Limit of Topological Insulator Thin Films of Sb$_2$Te$_3$. *Phys. Rev. Lett.* 2012, 108 (1), 016401.

(22) Wang, T.; Song, H.; He, K. Structural Design and Molecular Beam Epitaxy Growth of GaAs and InAs Heterostructures for High Mobility Two-Dimensional Electron Gas. *Quantum Front.* 2024, 3, 13.

(23) Yin, Y.; Wang, G.; Liu, C.; Huang, H.; Chen, J.; Liu, J.; Jia, J. et al. Moiré-Pattern-Modulated Electronic Structures in Sb$_2$Te$_3$/Graphene Heterostructure. *Nano Res.* 2022, 15, 1115–1119.

(24) Hsieh, D.; Qian, D.; Wray, L. A.; Xia, Y.; Hor, Y. S.; Cava, R. J.; Hasan, M. Z. Landau Levels in a Topological Insulator. *Phys. Rev. B* 2011, 84 (16), 161306.

(25) Fu, Y. S.; Hanaguri, T.; Igarashi, K.; Kawamura, M.; Bahramy, M. S.; Sasagawa, T. et al. Observation of

Zeeman Effect in Topological Surface State with Distinct Material Dependence. *Nat. Commun.* 2016, 7, 10829.

(26) Allison, G.; Mori, N.; Patanè, A.; Endicott, J.; Eaves, L.; Maude, D. K.; Hopkinson, M. Strong Effect of Resonant Impurities on Landau-Level Quantization. *Phys. Rev. Lett.* 2006, 96 (23), 236802.

(27) Okada, Y.; Zhou, W.; Dhital, C.; Walkup, D.; Ran, Y.; Wang, Z.; Madhavan, V. et al. Visualizing Landau Levels of Dirac Electrons in a One-Dimensional Potential. *Phys. Rev. Lett.* 2012, 109 (16), 166407.

(28) Ji, H.; Yuan, N. F. Q. Induced Zeeman effect of moiré surface states in topological insulators. arXiv Preprint arXiv:2507.04844 [cond-mat.mes-hall], 2025.